# Lifespan and propagation of information in On-line Social Networks: a Case Study

Giannis Haralabopoulos, Ioannis Anagnostopoulos

School of Sciences, Dpt of Computer Science and Biomedical Informatics

University of Thessaly Lamia, Greece

E-mail: {ghara, janag}@dib.uth.gr

Tel. +30 22310 66937

*Abstract-* Since 1950, information flows have been in the centre of scientific research. Up until internet penetration in the late 90s, these studies were based over traditional offline social networks. Several observations in "offline" information flows studies, such as two-step flow of communication and the importance of weak ties, were verified in several "online" studies, showing that the diffused information flows from one Online Social Network (OSN) to several others. Within that flow, information is shared to and reproduced by the users of each network. Furthermore, the original content is enhanced or weakened according to its topic, the dynamic and exposure of each OSNs. In such a concept, each OSN is considered a layer of information flows that interacts with each other. In this paper, we examine such flows in several social networks, as well as their diffusion and lifespan across multiple OSNs, in terms of user-generated content. Our results verify the perception of content and information connection in various OSNs.

I. INTRODUCTION

Information is constantly exchanged online. Between friends, acquaintances, family members, colleagues and even unknown individuals. The context varies; local or world news, general and scientific facts, quotes, personal preferences etc. This broad information exchange would not be possible without a communicative underlying environment, such as the Internet. Moreover, the creation of Online Social Networks (OSNs) and their adoption to our everyday lives, created new information-sharing schemes, where users can disseminate information quite fast and easily.

The information that flows in such schemes – along with their characteristics, properties and impact - has been the subject of several studies. Most of these studies are focused on virality, diffusion, dynamics, propagation and influence of information throughout a single OSN. However, nowadays, OSNs are densely connected to each other, through multiple information flows.

Considering every OSN as a layer (Fig. 1) and the information as links connecting each layer, we propose the concept of multilayer information flow. In this concept, information is spread from a source layer and propagates in multiple others. To test our proposal, we decided to focus on Reddit and its content.

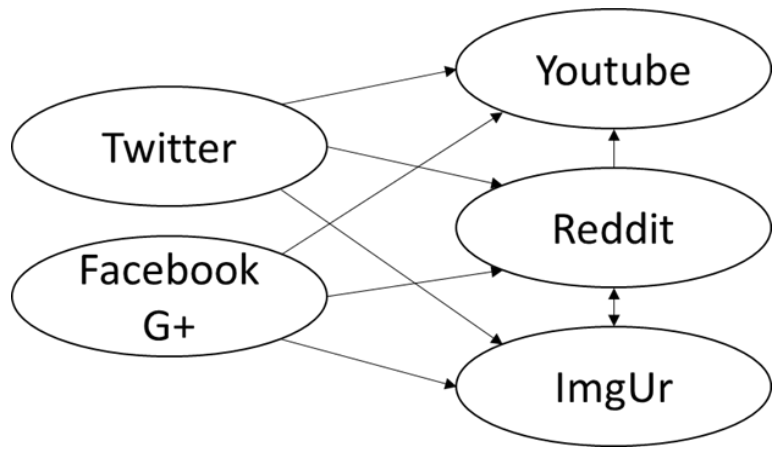

Figure 1. Linkage of different social domains

Reddit is a social news and entertainment site powered by user generated content. Registered users submit content on the form of a descriptive link that may contain; an image, meme, video, question, Ask Me Anything (AMA) session, and the community can then vote and comment on that post. In correspondence with votes, users that have created a post or commented one, gain or lose "karma", a Reddit-oriented metric for user ranking. Posts that acquire a high (vote ratio in a short period after their submission, are moved to the front page. It is apparent that Reddit community, defines the content and determine its "success" or "failure".

Across various content posted on Reddit, we focus on posts that link to an external domain and their traffic is measurable. Thus, content linking to Wikipedia articles or news sites, is not taken into consideration in this research. One of our initial observations, was that the highest rated content was mostly from the ImgUr domain. ImgUr is the most popular online image hosting service in Reddit community.

As illustrated in Figure 1, content on Reddit can be (amongst others) an image or a set of images hosted in ImgUr, or a video in YouTube. In ImgUr, content is usually created at the same time as the corresponding post in Reddit. While in YouTube, most posts in Reddit are linked to old videos. A short time after their creation and rise in terms of popularity within Reddit, users from different OSNs start mentioning that content, either by citing Reddit or the domain where the content is hosted (ImgUr or YouTube in our case).

For the purposes of our research, we wanted to use famous and heavily-visited OSNs. In one hand, Twitter provided us with the ability to fully observe the impact of a front-paged Reddit post. On the other hand, since most content in

Facebook and Google Plus is private, we only discovered a fraction of the total references, derived from search through public posts.

It is fairly obvious that information is shared and spread between these social networks. A post in any of these networks impacts the others as well. But, what is the size of this impact? How long it lasts? Is it dependent over different thematic categories? These are some of the question raised in this paper. User generated content coupled with multiple OSNs is a combinaton that can provide some really interesting results. Information flows, their propagation and diffusion along information virality and its topics are the main aspects addressed in our research. Through multiple social network analysis, we aim to present the perception of multi-layered flow of information in modern OSNs.

The remainder of the paper is as follows. In Section II we present the most important research initiatives along with their results over virality and information diffusion in social networks. Section III provides a detailed description in respect to our methodology. In Section IV we thoroughly present the results derived from our analysis, while in the last two sections (V and VI), we discuss evaluation issues and conclude our research.

## II. RELATED WORK

The topic of diffusion has been in the centre of sociology interest for many years. Even before the rise of Online Social Networks (OSNs), social ties and information flows have been studied in traditional real-life social networks.

The notion that information flows, from mass media to opinion leaders and later on to a wider population as final consumers, was firstly introduced during the middle of 40's [1]. In their introduction, Lazarsfeld et al., found that - during a presidential election - mass media had a direct influence on voting intentions, but informal and personal contact was more frequently mentioned as sources of influence. Almost a decade later, Kaltz and Lazarsfeld revisited the subject by proposing their theory of "Two step flow of communication" [2]. Specifically, they proposed that the information from media is firstly received by opinion leaders, who then pass their interpretation and actual information to other individuals. In a similar conceptualisation, Granovetter and Mark noted that analysis of social networks is suggested as a tool for linking micro and macro levels of sociology theory [3]. Small-scale interactions become large-scale patterns and feed back into small groups. The author concluded that weak ties are seen as the individual's opportunity towards integration into a community, while strong ties lead to overall fragmentation. Similar conclusions were verified in OSNs after nearly 40 years, while at the same time, the interest for social media analysis has been skyrocketed [4], [5].

Viral marketing was introduced in 1997 by Juvertson and Draper [6]. In their work, the authors described their marketing strategy for a free email service, in which by sending personal messages to individuals they boosted its user base. Their strategy proved to be very effective, gaining millions of users within few months. During the following decade, not only e-mail but several internet-based services evolved (e.g. real-time interactive services, OSNs), becoming an integral part of marketing strategies worldwide.

Porter and Golan in [7], found that provocative content such as sexuality, humour, violence and nudity are crucial virality factors, in comparison with traditional TV advertising, where emotive content had always been the key. In addition to virality, information diffusion became an important research subject. Leskovec et al. in [8], modelled outbreak detection via node selection, while performing a two-fold evaluation of their model by using a water distribution network as well as a blog network. Although the described model was not verified on any OSN, yet it could be easily applied in such networks, mainly because OSNs have similar diffusion properties with the ones tested in [8]. Concerning word of mouth scenarios, Allsop et al. in [9], note that 59% of individuals frequently share online content. The authors also tried to examine the link between emotion and virality, and concluded that content sharing is mainly adopted for entertainment purposes. Berger and Milkman in [10], observed that positive content is more viral than negative. Moreover, they ended up with the conclusion that the more the content evokes emotions of activation (e.g. anger, awe, anxiety), the more viral is, in contrast to deactivating emotions (e.g. softness).

The authors in [11] are focused on identifying the optimal network that best describes information propagation in news media and blogs. News diffusion networks were found to have a core periphery structure, where a small set of core media sites diffuses the information to the rest of the network, while blogs are mostly influenced by mass media. The authors of the work described in [12], analyse word of mouth through web means, and mainly through email forwarding. Two dynamic patterns were observed by the authors, namely "Transmissibility" and "Fanout Cohesion", while direct referrals proved the best form of diffusion, creating affinity paths.

Social influence modelling was the scope of [13]. Using old propagation data to create such models, the authors observed that viral marketing could leverage genuine influence, which -as they claimed- occurs only in real-world networks. Similarly, the authors in [14], investigate the social influence in meme social graph. They found out that most reports are made during the content creation period and shortly after it. Power users and post's age were proved crucial virality factors.

In [15], the authors study virality issues in Twitter. Due to its message-based nature, they observed that the direct mentions network, is of higher value, while the information chains formed were topic-related and presented a short time span. Finally, they noted that content lifespan is very important to virality, as also mentioned in [14]. In the same OSN, Sakaki et al. [16] study users as social sensors by monitoring information flow and dissemination during earthquake incidents. In their approach, they showed that earthquake detection can be equally effective compared to traditional monitoring techniques. In [17], authors analyse the entire Twitter graph in order to assess its topological characteristics. One of their findings was that each retweet would always reach 1000 users, regardless of its origin or content. Twitter network had low reciprocity, while at the same time a certain level of homophily was observed amongst its users.

In [18], Tucker measured the effectiveness of virality as a marketing tool. Unlike the work described in [6] -where direct emails messages were used- the author uses 400 videos and 24000 people to monitor information propagation. It was observed that for nearly every one million views, persuasiveness declined by 10%. However, this occurred only after a negativity threshold, above of which that effect is observed. That threshold was 6 million views reached only by 24 specific videos. In one hand, interaction affects were negative. On the other hand, sharing engaging, provocative (as defined in [7]) or humorous content and visual appeal, affected virality in a positively way.

In [19], the authors create a map of the so-called "life cycle" of the blog. They found out that blogs have a small cycle, and can be split into groups, based on their content and role. In their work, two of these groups, namely "elite" and "top political", not only create political information, but they also drive and sustain the viral process. Likewise, a virality study in OSNs conducted by Guerini et al., have shown that content is more important to virality than the influencer itself [20]. In this work, the authors used different metrics, such as buzz, appreciation, controversy and discussion raising. Similarly, focused in diffusion analysis of Twitter, Hansen et al. in [21], observed the effect of negativity in virality. More specifically, it was found that non-news positive sentiment and negative news are more likely to be retweeted, thus supporting overall virality. They conclude with the following: "If you want to be cited: Sweet talk to your friends or serve bad news to the public".

In other OSN-related researches, authors of [22] study the maximization of influence, based on the most influential users of a particular OSN. The work described in [23] examines the passivity of Twitter users. Moreover, the authors found out that most of users act as passive consumers, not forwarding news or content, thus high popularity does not imply high influence. They also argue that in order to maximize influence, one must overcome the passivity of the network. In [24] authors create an algorithm to describe influence propagation, in terms of likelihood, based on previous logs. They point out that sparsification is a fundamental data reduction operation.

Rajyalakshmi et al. in [25], propose a stochastic model for the diffusion of several topics. The authors discovered that strong ties play a significant role in virality, having homophily as a major contributor, as observed in [17]. Homophily stands as the tendency of individuals to bond with other individuals that have similar interests. Micro and macro scales of the network become relevant - similarly to [3] - and the authors note that, acts within groups have a global impact. Romero et al. in [26], analyse hashtag diffusion in Twitter and discover that initial adopters are fairly important, they also study two concepts within Twitter. The first one is "stickiness", which stands as the probability of adoption based on exposure, and the second is "persistence", which is defined as the continuing exposure to a hashtag. Their results validate the "complex contagion" principle of sociology.

On a different, but very interesting basis, Guerini et al. in [27] note that dynamics of Social Networks exist in scientific literature and use psycholinguistic analysis to determine abstract affect towards virality. The linguistic style proves an important factor and papers with easier to read abstracts are more frequently downloaded. Essentially, this links virality along with language usage.

Going back to OSNs, Huang et al. in [28], try to effectively select a number of nodes in order to monitor information diffusion. The authors ended up with the conclusion that degree centrality and out degree of a node, are very important. Additionally, they propose a new node centrality method to identify the monitoring capability of a node, namely monitoring centre.

Authors of [29], study the dynamics of the spreading process. Their research relies on the assumption that macroscopic level dynamics are explained by semantic, topological and temporal interactions in microscopic level. The proposed model is based on machine learning techniques, as well as on the inference of time dependent diffusion probabilities. Furthermore in [4], Bakshy et al., address the problem of information diffusion in OSNs, by performing a large-scale experiment of random exposement of information signals to 253 million users. The authors ended up with the statement -as seen in [3] and [17] that, strong ties are more influential but weak ties are responsible for propagating novel information.

Guille et al. in [30], address the issue of information adoption in Twitter, based on the assumption described in [29] in respect to micro and macro level dynamics. They present an adaptable graph-based prediction model that estimates and adapts its parameters using machine-learning techniques. Their results are similar with the ones described in [2].

In [31], the authors aim to answer why and how some messages in OSNs become viral. They model information ageing and competing message streams. Their obtained results imply a threshold above which a message becomes viral. Moreover, the authors note that competing streams further raise this threshold. Finally, Weng et al. in [32], analyse longitudinal micro blogging data. The authors highlight the importance of triadic closure and shortcuts based on traffic, regarding to the evolution of the social network. However, triadic closure is relevant only in the early stages of a user's lifetime, since after this stage, user linking is generated by the dynamic flow of network information.

## III. METHODOLOGY – DATASET DESCRIPTION

Our research is focused in Reddit, a social news and entertainment site. Content is generated and ranked by users, based on positive/negative (up/down respectively) votes. Newly created posts with a high enough[1] rank, reach the front page. The submitted content often links to an external domain and varies from simple news posts, political articles, Ask Me Anything sessions, up to entertaining pictures. The view count on their original source, is not always available (e.g. Wikipedia, News Sites), while it is also possible for a post to not be linked with any other external domain. Additionally, comments on posts also link to several other domains, yet this link relation is out of context of our research.

Since our initial idea was to be able to count viewership in various parallel social networks, the types of posts, hosted in domains without a view counter, were excluded from our analysis. Subsequently, we were able to monitor pictures in ImgUr and videos in YouTube, since in the front page of Reddit, there are always some posts that link to either one.

Generally, a post in Reddit consists of its title, content and comment section. Title part is self-explanatory, comment section is hosted within Reddit domain, while content is usually hosted in an external domain and rarely in Reddit. We focus on the hosting domain of a post, and more specifically on the content hosted on YouTube and ImgUr domains.

---

[1] http://amix.dk/blog/post/19588

We should note that our initial plan included the Quickmeme domain, which was a well-known meme creation site. Unfortunately, the use of Quickmeme in Reddit was banned in June of 2013.

In our work, we need domains that provide viewership counters in order to monitor and examine information propagation features. In one hand, ImgUr and YouTube count views according to the user's IP address, while on the other hand, Reddit only provides the voting count of a post. As such, the sum of negative and positive votes is used as the Reddit views counter. We were aware that both the absence of an IP based view counter and the method Reddit uses to fuse voting[1] introduces some inaccuracy in our results. However, since our main aim is towards propagation rather than viewership volume, such kind of inaccuracy does not affect our initial perception over multi-layer information flow.

In some preliminary tests we performed on Facebook and Google Plus, it was observed that both demonstrated a fairly low propagation, even on viral content. This happens because both APIs, do not provide access to private posts. Thus, only public posts were taken into consideration. Additionally, the network usage of Google Plus is fairly low compared to Facebook. Considering these factors, we decided to only include Facebook in our analysis. Furthermore in order to provide a more detailed picture of how information is spread amongst Facebook users, we not only take into account the post count, but also the "Like" count of each post. This method yields greater numerical results, but still relatively small compared to Twitter mentions. In Twitter we counted the mentions of either the Reddit post, or of the content in its respective domain. Finally, in order to present our results in a consistent format between different OSNs, we address all kinds of viewership count as "Units of Interest" (UoI). As such, a single Unit of Interest is equal to one content view in its domain (ImgUr or YouTube), one vote in Reddit, one mention in Twitter or one mention/like in Facebook.

During August and September of 2013, we scrapped two categories of Reddit (subreddits), named "new" and "rising", in an almost hourly-basis. The selection of these two categories was two-fold. Firstly, because we focus on propagation of newly created content, and secondly, because in both Twitter and Facebook APIs only recent posts can be accessed in their entirety. In both selected subreddits, content is refreshed every 2 minutes. During the 60 days of our scraping, almost 1 million posts were obtained and further analysed.

These posts are separated into several topics such as "pictures", "gaming", "funny", "news", "videos", "music", etc. As mentioned, not all of them were used in our research. Out of more than 950.000 posts in total, nearly 102.400 met our domain scraping criteria, belonging to either ImgUr or YouTube. Each post was analysed every one hour (in most cases in a less frequent rate), in order to measure the accumulated "Units of Interest" in their hosting domains, Twitter, Facebook and Reddit.

Additionally, in order to discard content that is not gaining attention rapidly, we employed a simple rule (mentioned hereafter as check criterion). In this rule, we examined if "Units of Interest" in Reddit are doubled (in absolute values) on every check, for the first 4 checks after the post creation. We found out that only a small percentage of total posts reached such high viewership counts within such a short time span. Specifically, only 0.66% of the tested posts passed this UoI check criterion among our dataset (682 out of nearly 102.400 posts).

## IV. RESULTS

In this section, we present the results derived from our scraping procedure and the respective analysis. Data is separated based on their topic, subreddit category and the domain they are hosted in. So "new" and "rising" in every chart denotes the corresponding (subreddit) category, while ImgUr and YouTube defines the hosting domain. Reddit topics of posts are "AdviceAnimals", "Aww", "Eathpon", "Funny", "Gaming", "Gifs", "Movies", "Music", "Pics", "TIL", "Videos" and "WTF".

### A. Posts and Counters

An important remark in our analysis was that "rising" category contained less posts linking to ImgUr and YouTube, compared to "new" category. This is perhaps due to the fact that more news and current events usually gain attention in a short timeframe, thus featured in "rising" subreddit. In total, we found nearly 45.000 ImgUr and YouTube posts in "rising" category and nearly 70.000 in "new" category. More specifically, the posts of the "rising" category revealed 40.966 links to ImgUr and 3.084 links to YouTube. Similarly, the distribution for posts of "new" category was 51.984 and 6.366 respectively.

We also observed that "new" subreddit content is more prone to become viral, compared to content in "rising" subreddit. Content featured in "new" subreddit was twice as likely to pass our UoI check, while content in "rising" category presented a higher initial UoI count, but fell short on our criterion (of at least double UoI). Out of 40.966 posts linking to ImgUr in "rising" subreddit, only 200 surpassed our check criterion (0.48%). As for posts linking to YouTube, out of 3.084 posts in total, only 14 went through our check (0.45%). On the other hand, in "new" subreddit, out of the

51.984 posts that linked to ImgUr, 431 doubled their UoI in 4 subsequent checks (0.82%). Similarly, the number of posts that linked to YouTube and successfully passed our UoI criterion, was 37 out of 6.366 (0.58%).

| Topic | Rising Category | | New Category | |
|---|---|---|---|---|
| | ImgUr | YouTube | ImgUr | YouTube |
| AdviceAnimals | 27 | 0 | 73 | 0 |
| Aww | 18 | 0 | 38 | 0 |
| Eathpon | 2 | 0 | 5 | 0 |
| Funny | 84 | 0 | 158 | 0 |
| Gaming | 25 | 0 | 69 | 0 |
| Gifs | 4 | 0 | 8 | 0 |
| Movies | 3 | 2 | 5 | 2 |
| Music | 0 | 1 | 0 | 3 |
| Pics | 29 | 0 | 56 | 0 |
| TIL | 0 | 1 | 0 | 1 |
| Videos | 0 | 10 | 0 | 29 |
| WTF | 8 | 0 | 19 | 2 |
| **Total** | **200** | **14** | **431** | **37** |

**Table 1.** Number of posts per topic (after the check criterion)

We must note here that our UoI criterion should not be considered as a virality validation, yet it provides strong evidence that the analysed posts have been viewed by enough viewers, within a small timeframe and the interest is not diminishing. Among 682 highly viewed posts, 631 were linked to ImgUr and 51 were linked to YouTube. The topics spread, according to the subreddit categories and the employed social platforms, are illustrated in Figure 2 and all respective values are depicted in Table 1.

Prior to discussing Table 1 and Figure 2, let us examine the topic titles. As mentioned in [9] and [10], positive and entertaining content was found as the most frequently shared topic. In our case, we could label "AdviceAnimals", "Funny" and "WTF topics", as entertaining content. "Aww" and "Eathpon" can be considered as containing emotive content, as described in [7]. "TIL" is mainly informative, while "Pics" and "Videos" characterise various content. Finally, "Gaming", "Movies" and "Music" contain user-centric and specific entertainment content.

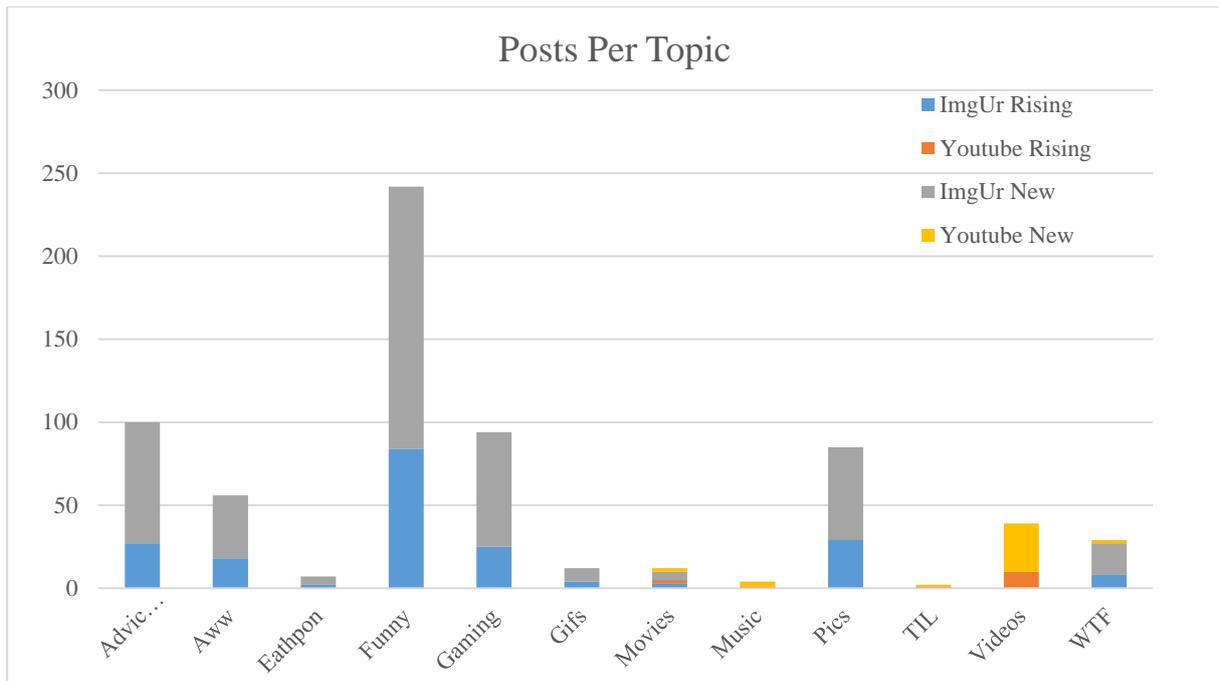

**Figure 2.** Distribution of posts per topic (after the check criterion)

As mentioned, entertaining and positive content is shared more frequently than anything else. Although provocative and controversial content can be created within Reddit and ImgUr, it rarely gets enough votes to appear in the front page. Most of the time, they appear in the form of a "news" or "TIL" post. The number of "gaming" posts that passed our check criterion is descriptive of the interests of Reddit users, especially when compared to other forms of established entertainment, such as movies and music.

| Topic | Rising Category | | | New Category | | |
|---|---|---|---|---|---|---|
| | Twitter and Facebook | Twitter | Facebook | Twitter and Facebook | Twitter | Facebook |
| AdviceAnimals | 5 | 8 | 0 | 11 | 29 | 0 |
| Aww | 3 | 11 | 0 | 4 | 32 | 0 |
| Eathpon | 0 | 2 | 0 | 0 | 5 | 0 |
| Funny | 21 | 51 | 0 | 47 | 87 | 0 |
| Gaming | 4 | 17 | 1 | 9 | 48 | 0 |
| Gifs | 0 | 4 | 0 | 1 | 7 | 0 |
| Movies | 2 | 3 | 0 | 3 | 3 | 0 |
| Music | 1 | 0 | 0 | 3 | 0 | 0 |
| Pics | 12 | 12 | 1 | 12 | 26 | 0 |
| TIL | 1 | 0 | 0 | 0 | 1 | 0 |
| Videos | 9 | 0 | 0 | 22 | 2 | 5 |
| WTF | 1 | 6 | 0 | 1 | 17 | 0 |
| **Total** | **59** | **114** | **2** | **113** | **257** | **5** |

**Table 2.** Number of posts shared in tested OSNs

Apart from the number of posts that passed our initial viewership test, we were strongly interested in their propagation to Twitter and Facebook. Out of the 682 posts, 371 were mentioned in Twitter, 7 in Facebook and 172 in both OSNs. Figure 3 and Table 2 show this distribution along with its respective values respectively.

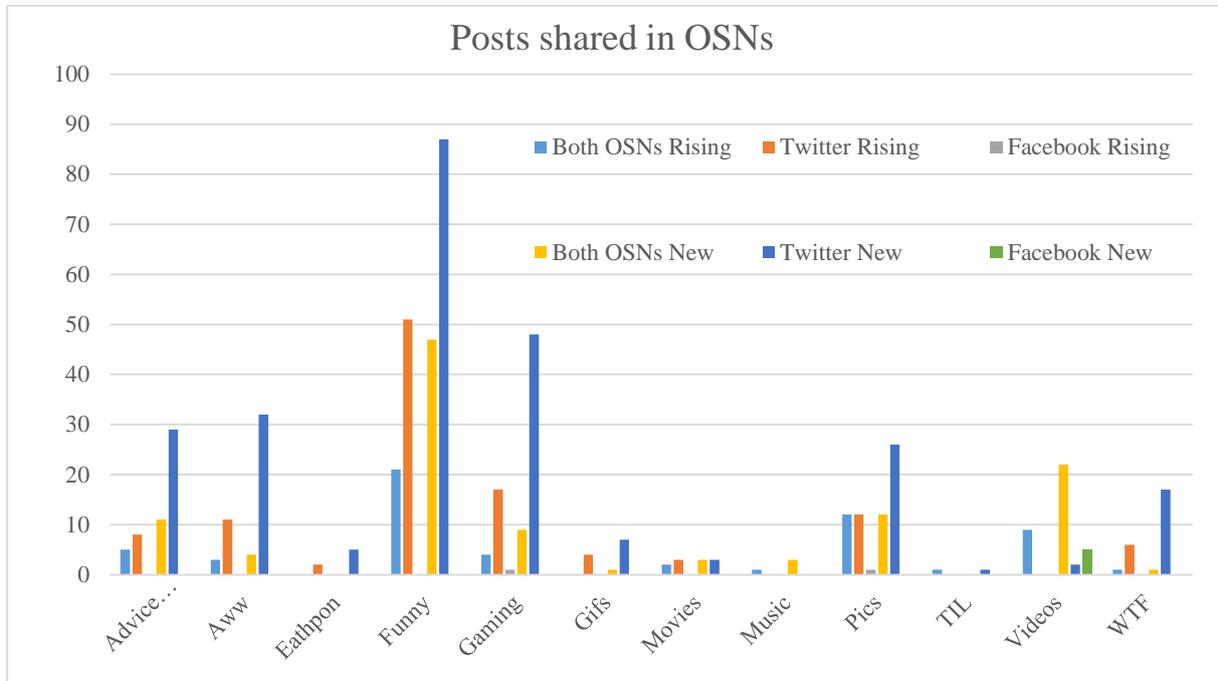

**Figure 3.** Distribution of posts that propagated to OSNs

In Figure 3, we can see that posts of "funny" topic, are the most shared in both Twitter and Facebook. However, a closer look in the percentage of posts shared in these OSNs, reveals that "funny" is not included among the top-shared topics (see Figure 4). Additionally, almost every category presents high Twitter and low Facebook sharing percentage (a direct result of the Facebook API limitations).

By further analysing the sharing percentage, we notice that almost all posts of "Music" and "TIL" categories are shared in both OSNs. However, the number of total viral posts in those categories is less than ten. The only categories that present a high number of viral posts propagating to other OSNs in a percentage level close to 50%, are "Videos" and "Movies". This confirms the positive attitude of internet users towards multimedia content. "Eathpon" posts, with photographs from around the world were 100% shared in Twitter.

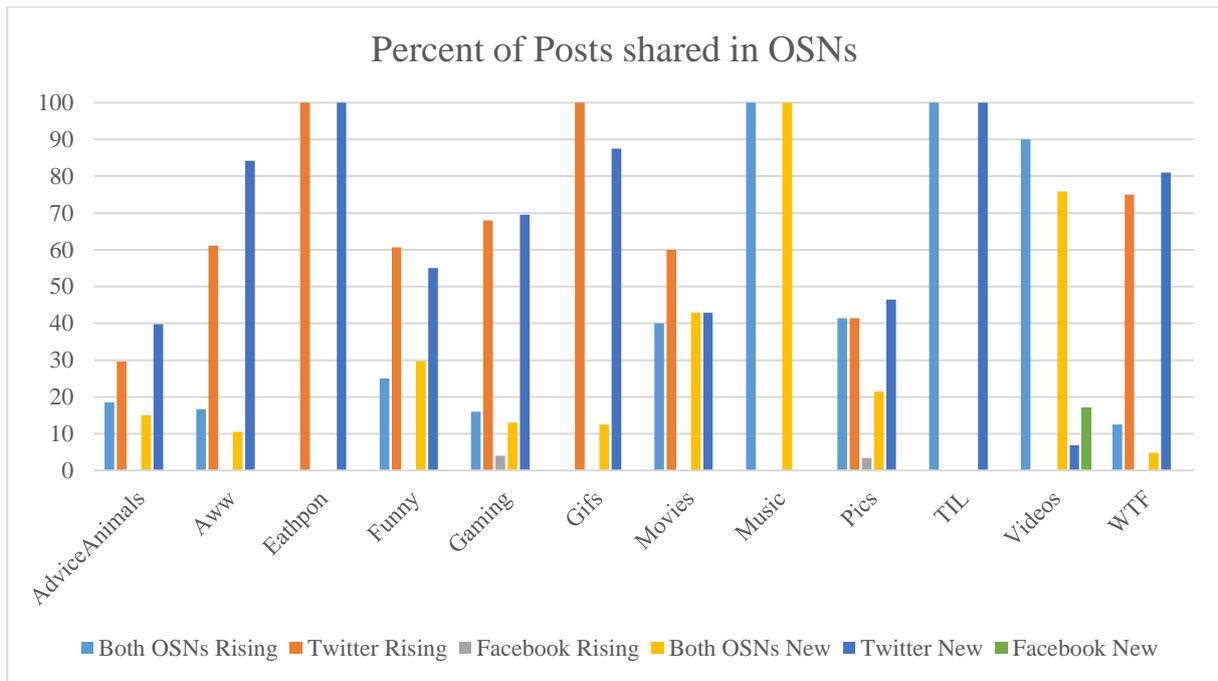

**Figure 4.** Percentage of posts that propagated to OSNs

As far as the topics with the highest number of posts are concerned ("AdviceAnimals", "Aww", "Funny", "Gaming" and "Pics"), we found out that the highest sharing percentage to both OSNs, pertained to "Pics" post from "rising" subreddit (41.38%), followed by "Funny" posts from "new" subreddit (29.75%). On the contrary, the lowest percentages appeared to "Aww" and "Gaming" posts from the "new" subreddit (10.53% and 13.04% respectively). Posts from "new" subreddit were more often shared in Twitter rather than posts from "rising" subreddit, for every topic except "Funny". Finally, it is worth noting the extremely low number of posts shared only in Facebook compared to those shared in both OSNs. Table 3 provides the percent values of posts propagating to OSNs. Sign (-) denotes zero propagation.

| Topic | Rising Category | | | New Category | | |
|---|---|---|---|---|---|---|
| | Twitter and Facebook | Twitter | Facebook | Twitter and Facebook | Twitter | Facebook |
| AdviceAnimals | 18,52 | 29,63 | - | 15,07 | 39,73 | - |
| Aww | 16,67 | 61,11 | - | 10,53 | 84,21 | - |
| Eathpon | - | 100 | - | - | 100 | - |
| Funny | 25 | 60,71 | - | 29,75 | 55,06 | - |
| Gaming | 16 | 68 | 4 | 13,04 | 69,57 | - |
| Gifs | - | 100 | - | 12,5 | 87,5 | - |
| Movies | 40 | 60 | - | 42,86 | 42,86 | - |
| Music | 100 | - | - | 100 | - | - |
| Pics | 41,38 | 41,38 | 3,45 | 21,43 | 46,43 | - |
| TIL | 100 | - | - | - | 100 | - |
| Videos | 90 | - | - | 75,86 | 6,9 | 17,24 |
| WTF | 12,5 | 75 | - | 4,76 | 80,95 | - |

**Table 3.** Percentage (%) of posts propagating to OSNs

## B. Online Social Network Propagation

As we already mentioned, UoI checks were simultaneously conducted on (at least) hourly-basis for Reddit, Twitter, Facebook and the domain the image or video was hosted (ImgUr or YouTube). Thus, we were able to observe the propagation of a certain Reddit post and content in both OSNs. We expected to measure a slow-rate information propagation from its source to both OSNs. In fact, such propagation pattern was indeed found and was the most common propagation standard throughout our analysed posts and topics..

For evaluation purposes we use the mean values for every topic and for each subreddit. When the mean value of OSN propagation is calculated, we do not take into consideration the posts that presented zero UoI variance in both OSNs. In other words, in our analysis we use the mean values of the posts that presented at least some variance in their UoI.

The time interval between subsequent checks performed in the tested domains (Reddit, ImgUr or YouTube, Twitter and Facebook) is at least one hour and the first check of every post has zero UoI. Each post is monitored for 7 days, yet we noticed that after 14 checks, the interest (in terms of UoI variation) declines significantly. The mean number of checks was 65 per post, while the $14^{th}$ check happens within a 24-hour period from the creation of the post.

Thus, in this section, we present the results obtained up to the first 14 checks, since after that check, we measured that the percentage level of UoI variation is usually lower than 0.1%. Moreover, we present the distinct observed patterns (separated per category) in X-Y axis, where X axis represents the check iteration and Y is the percentage UoI variance. Such a figure provides a sufficient visualization information in respect to the user interest per topic.

Furthermore, as we mentioned before, all posts are filtered to those that manage to double their UoI on Reddit in an hourly basis, for a 4 hour timeframe. Consequently, every post we analyse presents high Reddit UoI variance for the first 5 checks, but only few of them present higher levels of variance after that point. As far as our OSN data scraping is concerned, we only had access on information from the public API of Facebook, while on Twitter we can only check messages within a timespan of a week, due to Twitter API limitations.

The most common pattern observed is presented in Figure 5 (Pattern 1), and corresponds to "AdviceAnimals" posts from the ImgUr domain in the "rising" subreddit. Domain UoI are growing at a high rate, concurrently with Reddit.

Twitter is following with relatively small level of interest, while Facebook shows a delayed and low interest. Similar patterns appeared in several other categories, while in most of the cases Facebook presented nearly zero level of interest. This specific pattern was the most common in our research and appeared in 14 out of 18 ImgUr topics. Topics such as "AdviceAnimals", "Aww, "Funny", "Eathpon", "Gif" "Pics" and "WTF", followed this pattern - in both "rising" and "new" subreddits. However in both "Gif" categories, interest variation in Reddit is slightly higher than Domain.

This is a good evidence that Figure 5 (Pattern 1) models the most common propagation flow of Reddit content, to the tested OSNs. Content hosted in the parent Domain is visible from many external sources and rapidly accumulates interest, whereas Reddit posts grow in the same time with a lower UoI count variance. In our tested OSNs, firstly Twitter increases its -initially high- UoI, presenting low variance, in contrast with Facebook where its UoI grow later with lower initial numbers and with very low variance. More specifically, most Twitter links in Pattern 1 appeared simultaneously in numbers of 30 or more UoI, while Facebook usually starts with 10 or more UoI.

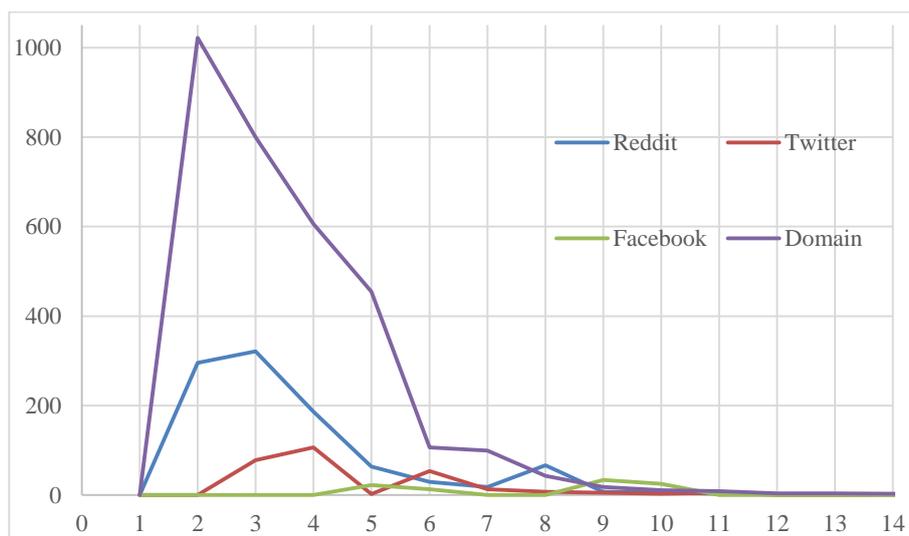

**Figure 5.** Pattern 1
Percent Variation of UoI through the first 14 checks
Topic: AdviceAnimals, Category: Rising, Domain: ImgUr

Another interesting pattern is illustrated in Figure 6 and was observed in "Movies" topic, in both subreddits. On every monitored post, we noticed interest spikes and, then sudden decreases in the UoI of all tested social media simultaneously, within few hours (Pattern 2). This is mainly because the content shared is usually time sensitive (e.g. new trailers or news for upcoming movies). In other words, such posts seem not to be so persistent, since they do not interest users for long periods, but they rapidly propagate to the tested OSNs. The concept of persistence in our analysis has to do with the ability of a post to retain interest for long periods of time, thus presenting positive UoI variation throughout that period.

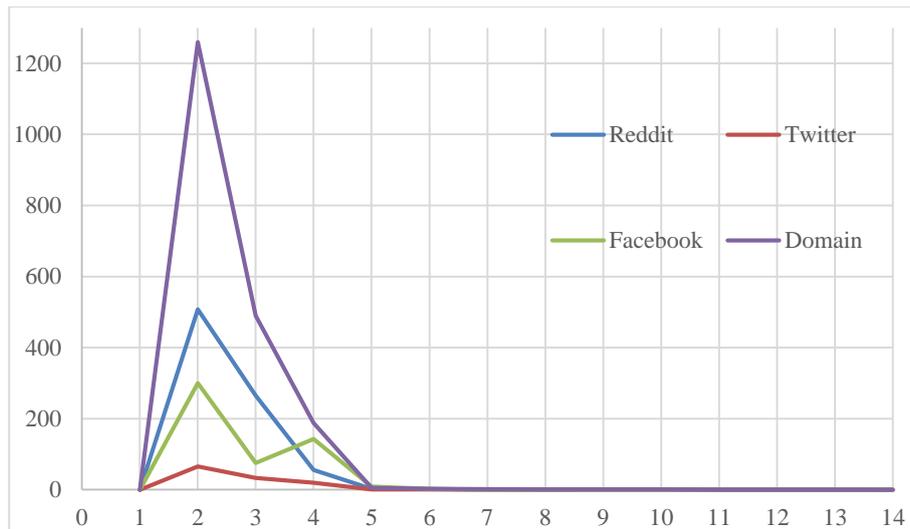

**Figure 6.** Pattern 2
Percent Variation of UoI through the first 14 checks
Topic: Movies, Category: New, Domain: ImgUr

We should add that "Movies" and "Gaming" posts, as dealing with two of the most popular forms of entertainment, present the highest propagation levels through OSNs. Posts about movies are generally propagating faster, but gaming posts appeared to be more persistent. This is highlighted in Figure 7 (Pattern 3), where Reddit interest intensifies with rates similar to Domain. Moreover, interest in ImgUr, Reddit and Twitter persists over the 10th check, which is a very unique case, since usually in all other posts, the UoI variance is practically at zero levels long before that point.

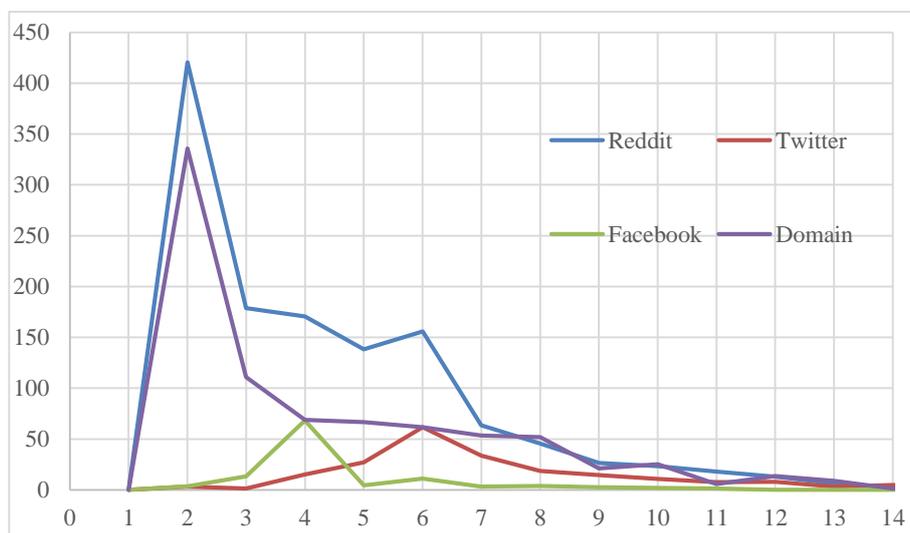

**Figure 7.** Pattern 3
Percent Variation of UoI through the first 14 checks
Topic: Gaming, Category: Rising, Domain: ImgUr

Interestingly enough, in YouTube content, we encountered completely different propagation patterns (Figure 8). In most cases, domain UoI were not seriously affected by the increased interest in OSNs and Reddit. This is mainly because shared content in YouTube is not usually a newly created video, but an instance of a video that existed for a long time and becomes mildly popular in time (Pattern 4). In our evaluation, the videos under study presented a very high initial UoI count, which resulted in a low UoI variance. Apart from that, we observed high UoI variance in Reddit and Twitter, low UoI variance in Facebook, and practically zero UoI variance in the tested domain.

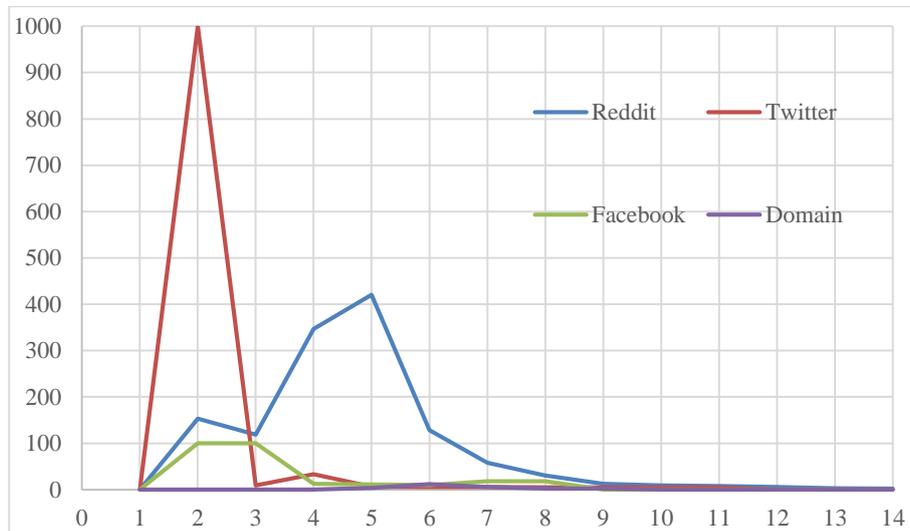

**Figure 8.** Pattern 4
Percent Variation of UoI through the first 14 checks
Topic: TIL, Category: Rising, Domain: YouTube

The pattern of Figure 9 (Pattern 5), presents what happens when content is created in YouTube at the same time with a post in topic "Videos" of "new" subreddit. Reddit, Twitter, Facebook and Domain UoI are increasing simultaneously, at the second check. However, in the next check, the domain interest is increased, while the interest in each of the OSNs and Reddit seems to rapidly decrease. Surprisingly, the high domain interest in the content leads to increased interest in Reddit and Twitter after 3 to 5 check intervals. Public Facebook interest appears after the spikes in Reddit and Twitter, maintaining the slower reaction which was encountered in almost every topic and subreddit.

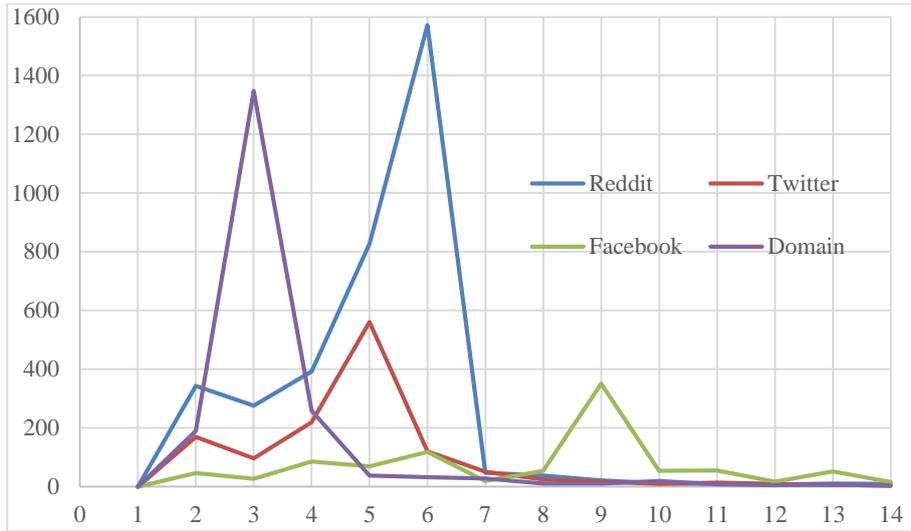

**Figure 9.** Pattern 5
Percent Variation of UoI through the first 14 checks
Topic: Videos, Category: New, Domain: YouTube

Table 4 depicts all five clustered patterns according to Reddit content topics (Topic), subreddit category (Category) and hosting domain (Domain). By analysing our scraped data, we noticed that Patterns 1, 2, and 3 reflect to ImgUr hosting domain, while Patterns 4 and 5 to YouTube. In bold we denote the representative patterns as presented in Figures 5 to 9, below them we have listed every topic that has a similar or identical pattern.. In respect to Table 4, every discovered pattern is illustrated in the Appendix.

| Discovered Patterns ( Topic – Category) | | | | |
|---|---|---|---|---|
| Pattern 1 | Pattern 2 | Pattern 3 | Pattern 4 | Pattern 5 |
| AdviceAnimals Rising<br>Aww Rising<br>Eathpon Rising<br>Funny Rising<br>Gifs Rising<br>Pics Rising<br>WTF Rising<br>AdviceAnimals New<br>Aww New<br>Eathpon New<br>Gifs New<br>Funny New<br>Pics Rising<br>WTF New | Movies Rising<br>Movies New | Gaming Rising<br>Gaming New | TIL Rising<br>TIL New<br>WTF New<br>Movies New | Videos Rising<br>Music Rising<br>Movies Rising<br>Videos New<br>Music New |
| Domain: ImgUr | | | Domain: YouTube | |

**Table 4.** Discovered Pattern per Topic

## V. DISCUSSION

Our analysis verified many observations of previous researches, mainly in respect to the micro and macro effects of information flows [1], [2], [29]. More specifically, a single post (micro effect) in the parent domain connected with a post in Reddit, starts to accumulate views (macro effect) up to a point where the information hops to OSNs (first in Twitter and then in Facebook) flowing through individuals, while eventually the interest dies off. Positive content is confirmed to be the mostly shared content in OSNs, as mentioned in [7] and [10]. Similarly to [14], we found out that a Reddit post is mostly popular within the first hours after its creation. This effect is also enhanced by the classification method of Reddit, where new posts need fewer votes (compared to old posts) to move to the first page.

In addition, we have observed that the most shared posts are of positive and entertaining content, also seen in [9] and [10]. These categories, would include post from topics such as "AdviceAnimals", "Funny" and "WTF topics, while we also identified emotive posts, as described in [7], in topics such as "Aww" and "Eathpon". Unfortunately, not enough viral posts of such content were found, in order to safely identify their propagation model. "Gaming", "Movies" and "Music" contain user-centric and very specific entertainment content, with low post appearance but great viral to total posts ratio, while, "TIL" is mainly informative and rarely used in posts linking to either ImgUr or Youtube.

Figure 10 illustrates the information flow scheme of the posts propagated to OSNs, according to our proposed method. Content creation in ImgUr is always tied to the respective post in Reddit occurring in $t_0$. Moreover, after approximately 3 hours, links to ImgUr or Reddit are shared in Twitter. Finally, after nearly12 hours, the same links appear in public posts of Facebook. This results in the following user interest distribution: for each Facebook view or like, we have 8 Twitter posts, 19.231 Reddit votes and 426.656 views on ImgUr, which is the hosting domain. This reveals that ImgUr UoI are 22 times the UoI of Reddit. This is due to the fact that not every user votes on a post and we only count up and down votes of posts as user actions similar to views. Additionally, another factor is the large number of external sources that link directly to the hosting domain. Furthermore, total Reddit votes were found to be 2.403 more than the number of Twitter mentions, which is significantly lower than Domain to Reddit views ratio. Although these are mean values, views ratio (Domain to Reddit, Reddit to Twitter etc.) dispersion was fairly low. Figure 11 presents the mean UoI values -in logarithmic scale- per Reddit post, in respect to the first 14 checks for ImgUr, while Table 5 depicts all respective values.

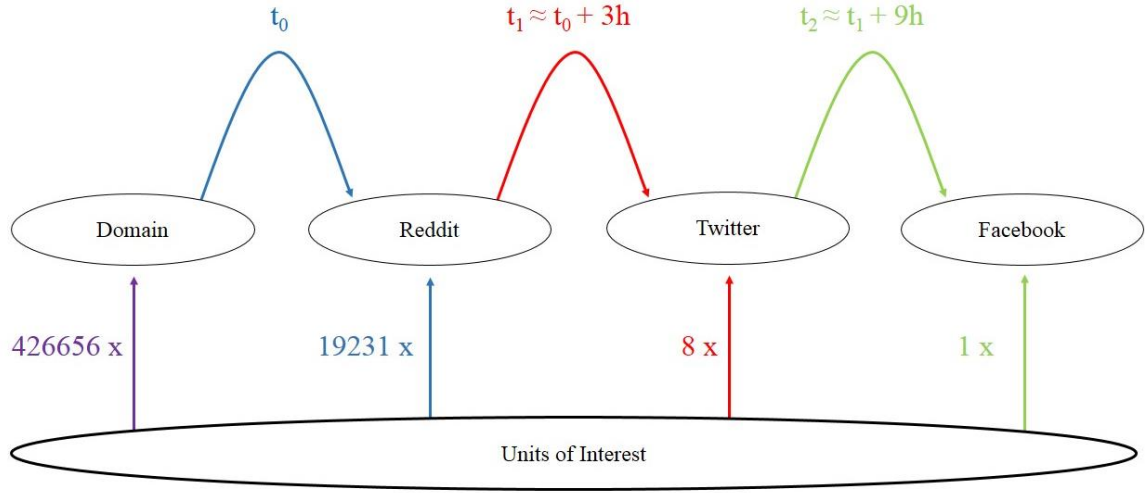

**Figure 10.** Propagation time and UoI allotment for ImgUr content of both subreddits

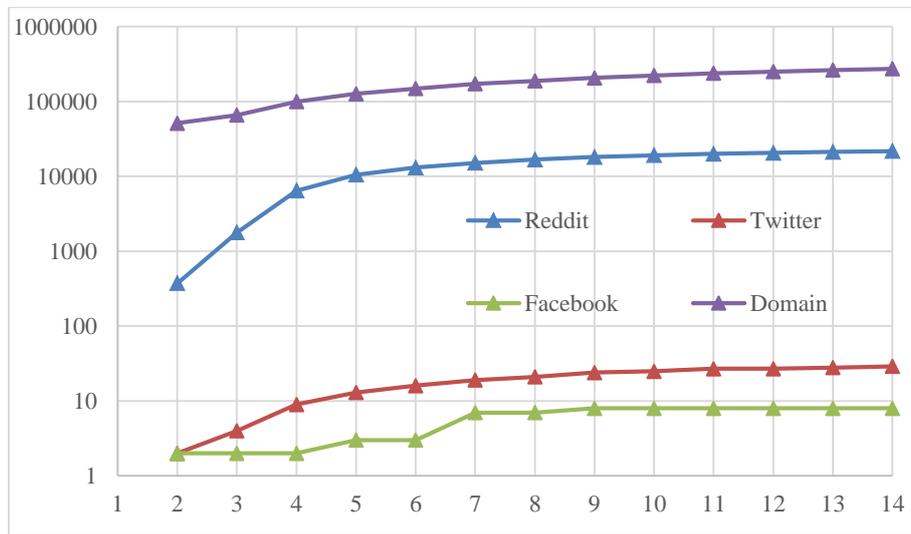

**Figure 11.** Mean UoI (in log scale) per Reddit post for the first 14 checks

Domain: ImgUr

| Check (#) | 1st | 2nd | 3rd | 4th | 5th | 6th | 7th |
|---|---|---|---|---|---|---|---|
| Reddit | - | 374 | 1.781 | 6.430 | 10.522 | 13.114 | 15.145 |
| Twitter | - | 2 | 4 | 9 | 13 | 16 | 19 |
| Facebook | - | 2 | 2 | 2 | 3 | 3 | 7 |
| ImgUr | - | 51.509 | 65.854 | 99.575 | 127.118 | 149.445 | 172.429 |
| Check (#) | 8th | 9th | 10th | 11th | 12th | 13th | 14th |
| Reddit | 16.750 | 18.136 | 19.149 | 19.971 | 20.618 | 21.213 | 21.721 |
| Twitter | 21 | 24 | 25 | 27 | 27 | 28 | 29 |
| Facebook | 7 | 8 | 8 | 8 | 8 | 8 | 8 |
| ImgUr | 189.017 | 207.424 | 223.305 | 238.900 | 250.917 | 263.924 | 273.495 |

**Table 5.** Mean UoI numerical values for the first 14 checks, Domain: ImgUr

Similarly, Figure 12 illustrates the respective information flow for the YouTube case. However, $t_0$ does not correspond to the time of creation in YouTube, but instead is the time at which the post in Reddit appears. In one hand, we found out that Reddit posts with content linking in YouTube, are rarely created at the same time as the video itself, since the video achieves a high number of viewers, far after its initial "viral" period. On the other hand, the first post in Twitter and Facebook appears much faster (approximately 2 and 3 hours after the Reddit post respectively). For each Facebook view or like, we measured 9 Twitter mentions, 2.162 Reddit votes and almost 1.7 million YouTube views. This difference is because most videos, when used in Reddit posts, are already widely known. Figure 13 presents the mean UoI values -in logarithmic scale- per Reddit post, in respect to the first 14 checks for YouTube, while Table 6 depicts all respective values.

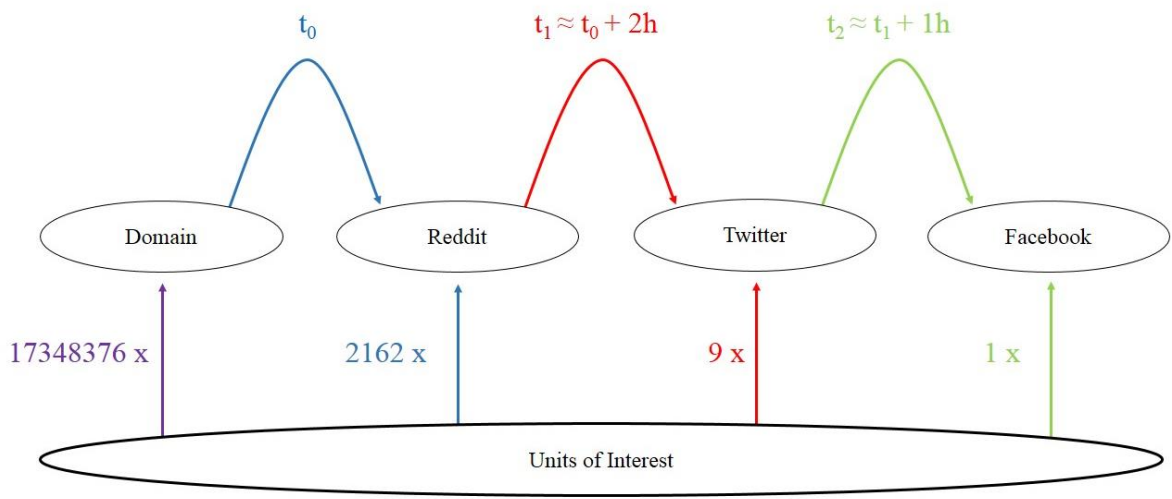

**Figure 12.** Propagation time and UoI allotment for YouTube content of both subreddits

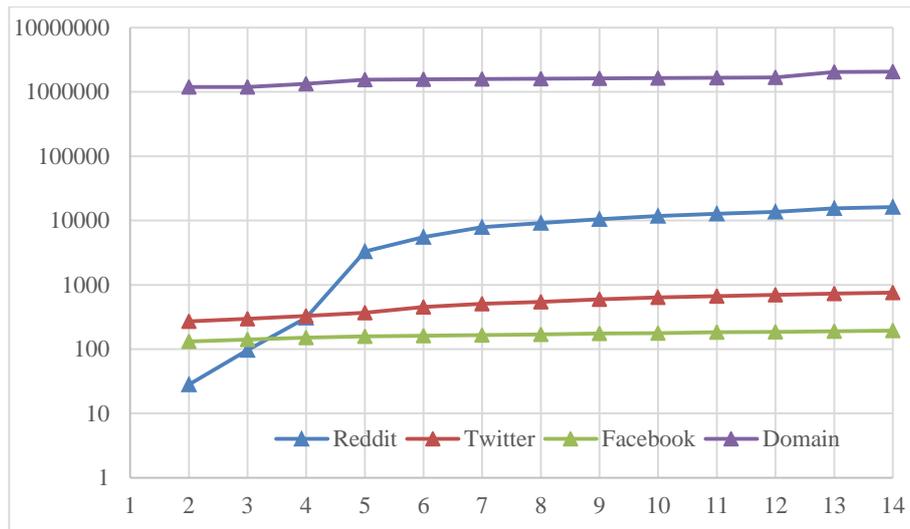

**Figure 13.** Mean UoI (in log scale) per Reddit post for the first 14 checks

Domain: YouTube

| Check (#) | 1st | 2nd | 3rd | 4th | 5th | 6th | 7th |
|---|---|---|---|---|---|---|---|
| Reddit | - | 28 | 97 | 305 | 3.321 | 5.554 | 7.897 |
| Twitter | - | 269 | 296 | 327 | 368 | 454 | 505 |
| Facebook | - | 132 | 141 | 150 | 158 | 162 | 165 |
| ImgUr | - | 1.197.409 | 1.197.785 | 1.329.408 | 1.546.500 | 1.566.521 | 1.586.154 |
| Check (#) | 8th | 9th | 10th | 11th | 12th | 13th | 14th |
| Reddit | 9.133 | 10.489 | 11.695 | 12.785 | 13.665 | 15.389 | 16.120 |
| Twitter | 544 | 594 | 637 | 665 | 696 | 730 | 755 |
| Facebook | 170 | 175 | 177 | 183 | 186 | 189 | 195 |
| ImgUr | 1.607.647 | 1.613.988 | 1.643.021 | 1.669.383 | 1.680.082 | 2.048.028 | 2.070.141 |

**Table 6.** Mean UoI numerical values for the first 14 checks, Domain: YouTube

Of course, these figures highlight the analysis of data from our tested domains and OSNs, during our experimentations. However, if we used a slightly different method, results would be significantly different. For example, if we would calculate the exposure on both OSNs, UoI allotment would be smoother. According to "An Exhaustive Study of Twitter Users Across the World"[2], the mean number of followers in an average Twitter user has been estimated equal to 208. That would mean that Reddit votes and Twitter mentions having a ratio of almost 1 in YouTube content and 0.1 in ImgUr content (down from 240 and 2.403, respectively). Furthermore, if we add the mean number of friends in an average Facebook account, which according to the "Anatomy of Facebook"[3] is equal to 190, we would get slightly higher Twitter to Facebook ratio, but significantly lower Reddit to Facebook and Domain to Facebook ratios. Finally, the fact that only public Facebook can be scraped, is a big factor to those particular UoI allotments.

---

[2] http://www.beevolve.com/twitter-statistics/, published results from October 2012

[3] https://www.facebook.com/notes/facebook-data-team/anatomy-of-facebook/10150388519243859

## VI. CONCLUSIONS – FUTURE WORK

Information starts within a domain and hops to various other media within minutes. However, upon its propagation, information flow within the domain does not halt - it just slows down. Entertainment and positively emotive content is the most "viral". Persistence was only found in gaming posts, while posts with movie content were the only ones that spread nearly simultaneously to every domain and OSN.

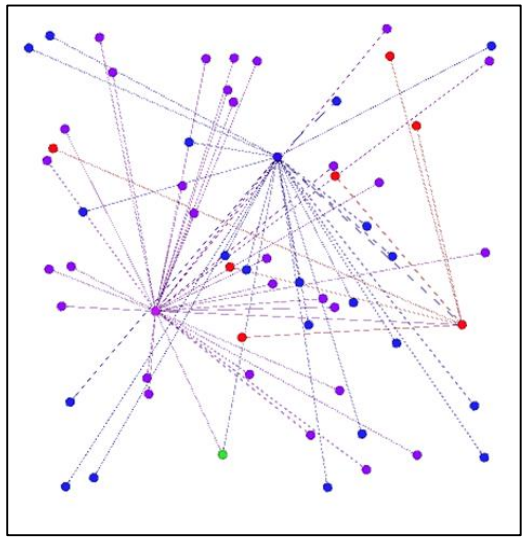

**Figure 12.** Information flow graph

We should consider a new perception of information flow, one that would accurately describe content sharing in modern OSNs. In order to provide a simple example of such information sharing flow, let us consider a random Reddit post that follows Pattern 1 (as presented in Figure 5), which has 30 domain UoI, 20 Reddit UoI, 7 Twitter UoI and 1 Facebook UoI. The UoI count used in this post, is obtained right after its first Facebook UoI appears, long before the $14^{th}$ check. As illustrated in Figure 12, a single thread of information is spread into multiple concurrent users, spanning to various users of different OSNs simultaneously. Then a simple clustering, reveals the perception of layers (Figure 13). Information flows from the top layer towards the lower one, concurrently with the vertical flow, where each layer expands. After approximately 12 hours, and according to Pattern 1, graph expansion halts.

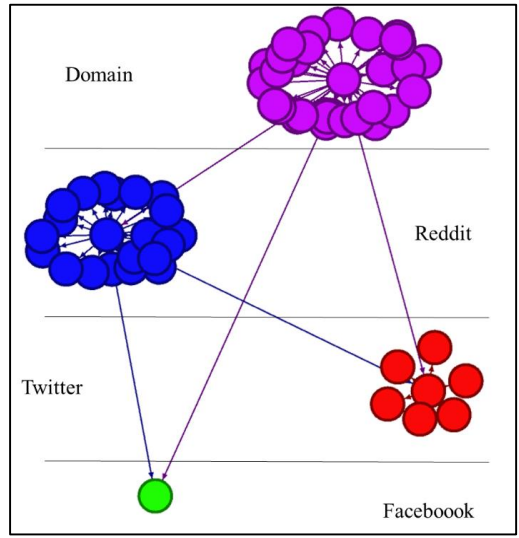

**Figure 13.** Information flow through layers

Subsequently, we look forward into analysing a wider range of topics, in as many OSNs as we can. We believe that a predictive algorithm for viral posts will greatly benefit from our scraping findings. Furthermore, we observed that various news posts in Reddit, have arisen to the first page before getting reported in dedicated news sites. As such, we could consider a different perspective, where posts could serve as a worldwide events station, similarly to what is proposed in [17]. Although we didn't discuss the great personalization options and properties of Reddit, we look forward into researching all these aspects of this modern OSN and its information flows in a future work of ours.

**REFERENCES**

1. P. F. Lazarsfeld, B. Berelson, H. Gaudet, The People's Choice: How the Voter Makes up his Mind in a Presidential Campaign, Columbia University Press, 1944

2. E. Katz and P. F. Lazarsfeld, Personal Influence, The Part Played by People in the Flow of Mass Communications. Transaction Publishers, 1970,

3. Granovetter and Mark S., The Strength of Weak Ties, American Journal of Sociology, 1973, pp. 1360-1380

4. E. Bakshy, R. Itamar, M. Cameron, and L. Adamic, The Role of Social Networks in Information Diffusion, Proc. of the 21st International Conf. on World Wide Web, 2012, pp. 519-528

5. S. Rajyalakshmi, B. Amitabha, D. Soham and R. M. Tripathy, Topic Diffusion and Emergence of Virality in Social Networks, arXiv preprint arXiv:1202.2215, 2012

6. S. Jurvetson, and T. Draper, Viral Marketing: Viral Marketing Phenomenon Explained, DFJ Network News, 1997

7. L. Porter and G. J. Golan, From Subservient Chickens to Brawny Men: A Comparison of Viral Advertising to Television Advertising." Journal of Interactive Advertising 6 (no. 2), 2006, pp. 30-38

8. J. Leskovec, A. Krause, C. Guestrin, C. Faloutsos, J. VanBriesen and N. Glance, Cost-Effective Outbreak Detection in Networks, Proc. of the 13th ACM SIGKDD international Conf. on Knowledge Discovery and Data Mining, 2007, pp. 420-429

9. D. T. Allsop, B. R. Bassett and J. A. Hoskins, Word-of-Mouth Research: Principles and Applications, Journal of Advertising Research 47 (no. 4), p. 398, 2007

10. J. Berger and K. Milkman, Social Transmission, Emotion, and the Virality of Online Content, Wharton Research Paper, 2010

11. M. G. Rodriguez, J. Leskovec and A. Krause, Inferring Networks of Diffusion and Influence, Proc. of the 16th ACM SIGKDD International Conf. on Knowledge Discovery and Data mining, pp. 1019-1028, 2010

12. J. L. Iribarren, and E. Moro, Affinity Paths and Information Diffusion in Social Networks, Social Networks 33 (no. 2), pp. 134-142, 2011

13. A. Goyal, F. Bonchi and L. VS Lakshmanan, Learning Influence Probabilities in Social Networks, Proc. of the 3rd ACM International Conf. on Web Search and Data Mining, pp. 241-250, 2010

14. D. Ienco, F. Bonchi and C. Castillo, The Meme Ranking Problem: Maximizing Microblogging Virality, Data Mining Workshops, pp. 328-335, 2010

15. J. Yang and S. Counts, Predicting the Speed, Scale, and Range of Information Diffusion in Twitter, ICWSM, pp. 355-358, 2010

16. T. Sakaki, M. Okazaki and Y. Matsuo, Earthquake Shakes Twitter Users: Real-Time Event Detection by Social Sensors, Proc. of the 19th International Conf. on World Wide Web, pp. 851-860, 2010

17. H. Kwak, C. Lee, H. Park and S. Moon, What is Twitter, a Social Network or a News Media?, Proc. of the 19th International Conf. . on World Wide Web, pp. 591-600, 2010

18. C. Tucker, Ad Virality and Ad Persuasiveness, Available at SSRN 1952746, 2011

19. K. Nahon, J. Hemsley, S. Walker and M. Hussain, Blogs: Spinning a Web of Virality, Proc. of the iConference, pp. 348-355, 2011

20. M. Guerini, C. Strapparava, and G. Özbal, Exploring Text Virality in Social Networks, ICWSM, 2011

21. L. K. Hansen, A. Arvidsson, F. Å. Nielsen, E. Colleoni and M. Etter, Good Friends, Bad News-Affect and Virality in Twitter, Future Information Technology, pp. 34-43, 2011

22. F. Bonchi, Influence Propagation in Social Networks: A Data Mining Perspective, IEEE Intelligent Informatics Bulletin 12 (no. 1), pp. 8-16, 2011

23. D. M. Romero, W. Galuba, S. Asur and B. A. Huberman, Influence and Passivity in Social Media, Machine Learning and Knowledge Discovery in Databases, pp. 18-33, 2011

24. M. Mathioudakis, F. Bonchi, C. Castillo, A. Gionis and A. Ukkonen, Sparsification of Influence networks, Proc. of the 17th ACM SIGKDD International Conf. on Knowledge Discovery and Data Mining, pp. 529-537, 2011

25. S. Rajyalakshmi, A. Bagchi, S. Das and R. M. Tripathy, Topic Diffusion and Emergence of Virality in Social Networks, arXiv preprint arXiv:1202.2215, 2012

26. D. M. Romero, B. Meeder and J. Kleinberg, Differences in the Mechanics of Information Diffusion Across Topics: Idioms, Political Hashtags, and Complex Contagion on Twitter, Proc. of the 20th International Conf. on World Wide Web, pp. 695-704, 2011

27. M. Guerini, A. Pepe and B. Lepri, Do Linguistic Style and Readability of Scientific Abstracts Affect their Virality?, ICWSM, 2012

28. Y. Li, S. Huang, C. Fan and G. Yang, The Selection of Information Diffusion Monitoring Nodes in Directed Online Social Networks, Proc. of the 2012 International Conf. on Information Technology and Software Engineering, pp. 533-540, 2013

29. A. Guille, and H. Hacid, A Predictive Model for the Temporal Dynamics of Information Diffusion in Online Social Networks, Proc. of the 21st International Conf. Companion on World Wide Web, pp. 1145-1152, 2012

30. A. Guille, H. Hacid and C. Favre, Predicting the Temporal Dynamics of Information Diffusion in Social Networks, arXiv preprint arXiv:1302.5235, 2013

31. A. Karnika, A. Saroopb and V. Borkarc, On the Diffusion of Messages in On-line Social Networks, Performance Evaluation Volume 70 (Issue 4), pp. 271–285, 2013

32. L. Weng, J. Ratkiewicz, N. Perra, B. Goncalves, C. Castillo, F. Bonchi, R. Schifanella, F. Menczer and A. Flammini. Proc. of the 19th ACM SIGKDD Conf. on Knowledge Discovery and Data Mining, 2013

# APPENDIX

## PATTERN 1: TOPIC / CATEGORY / DOMAIN: IMGUR

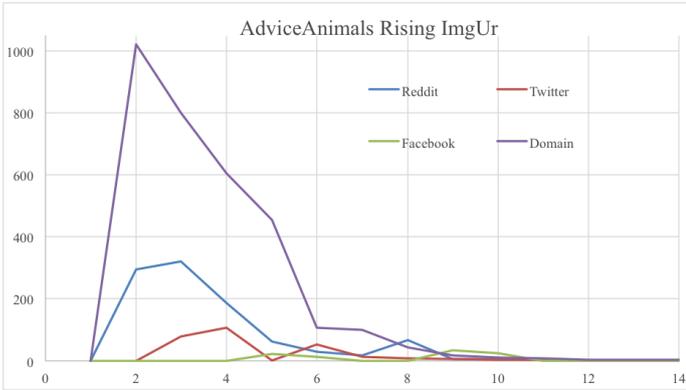

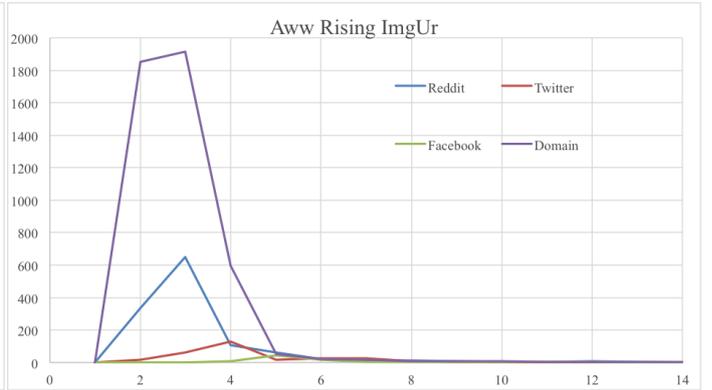

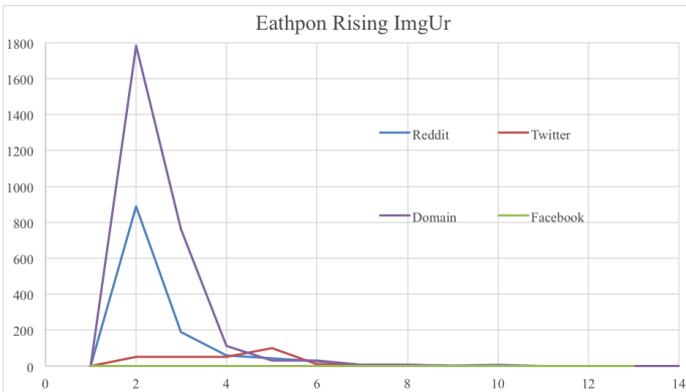

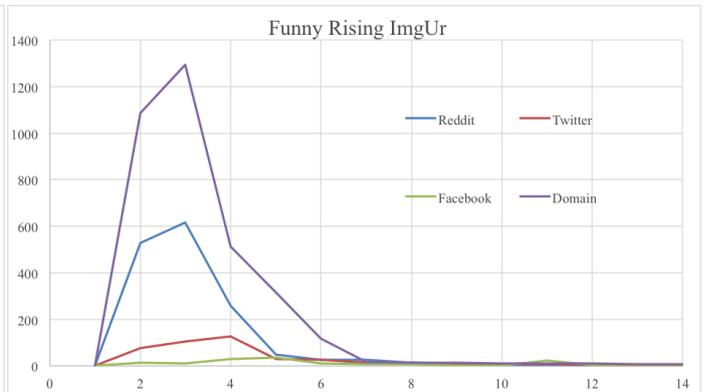

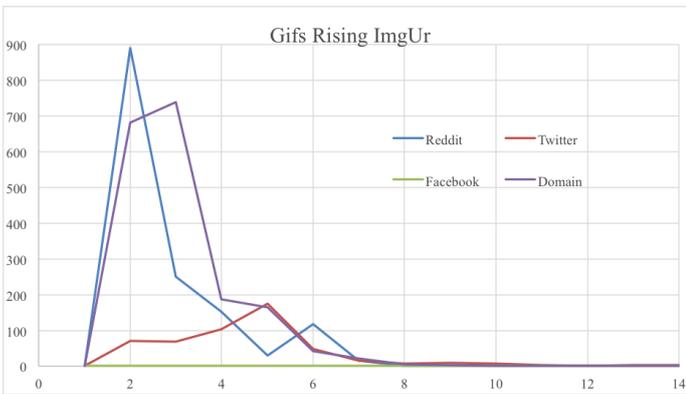

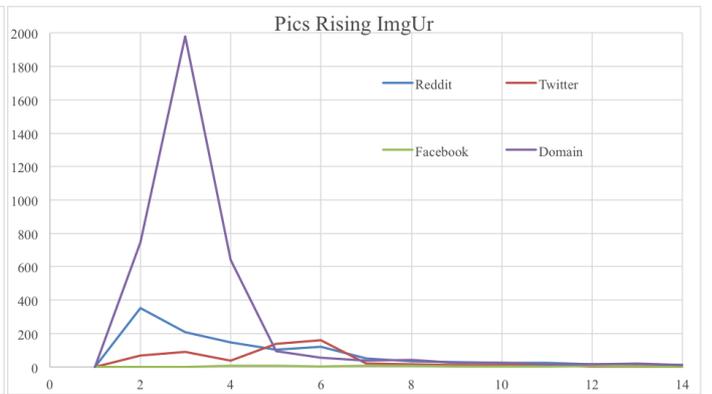

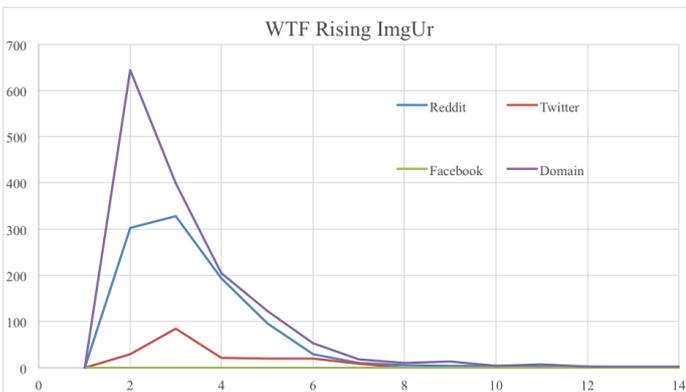

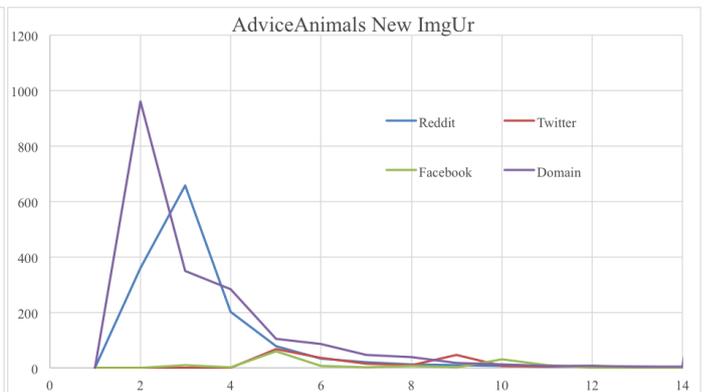

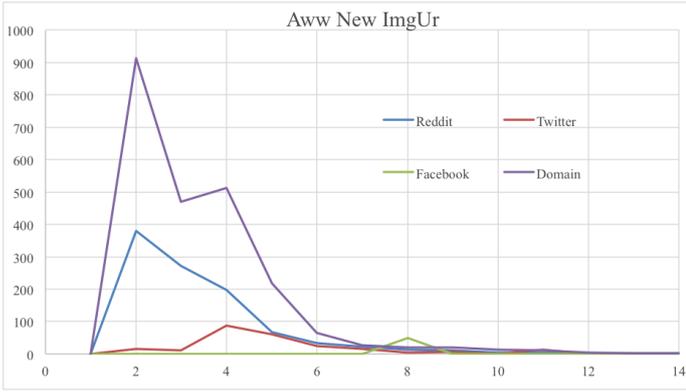
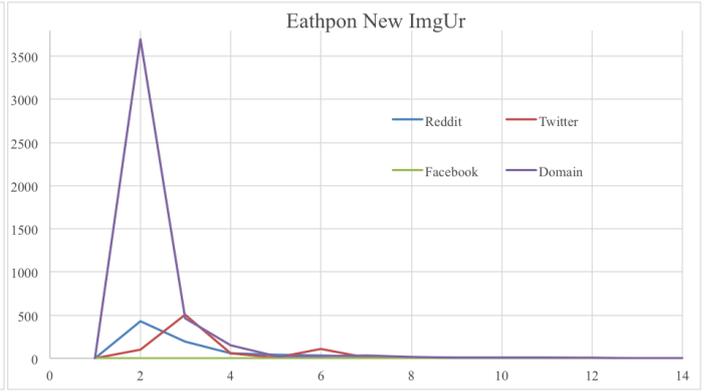
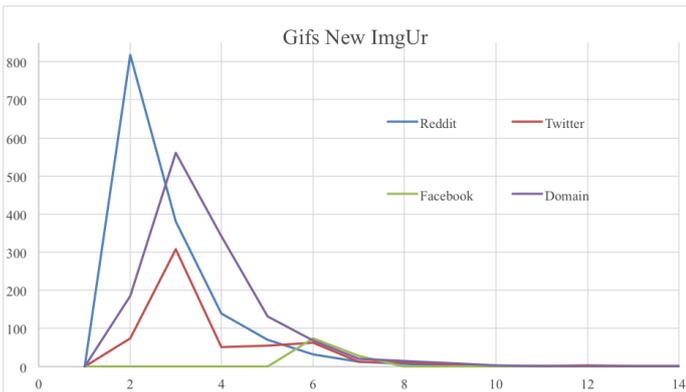
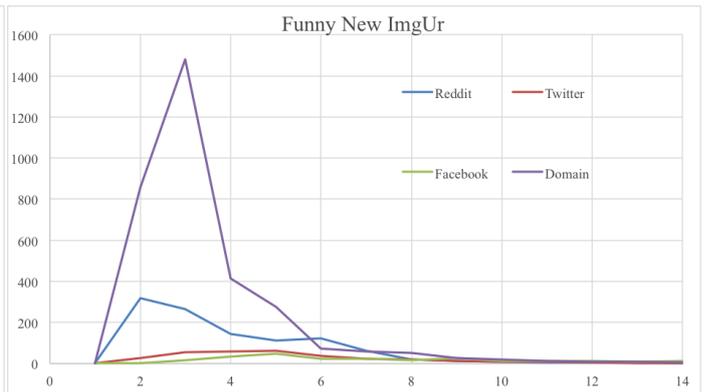
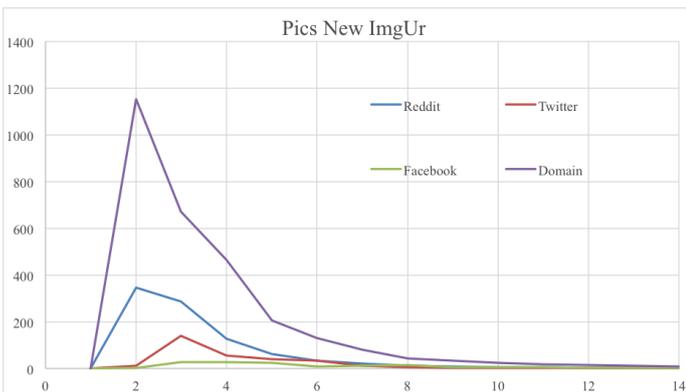
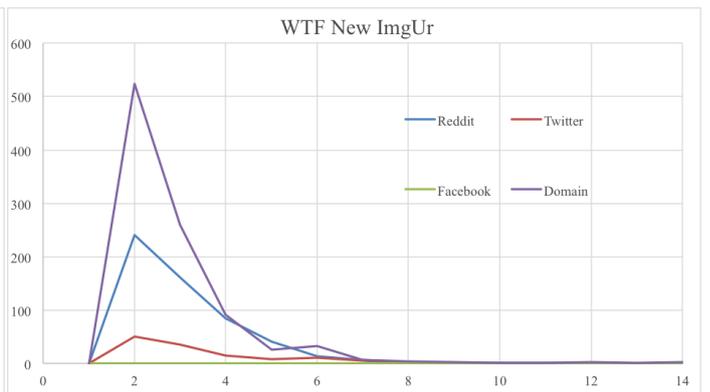

**PATTERN 2: Topic / Category / Domain: ImgUR**

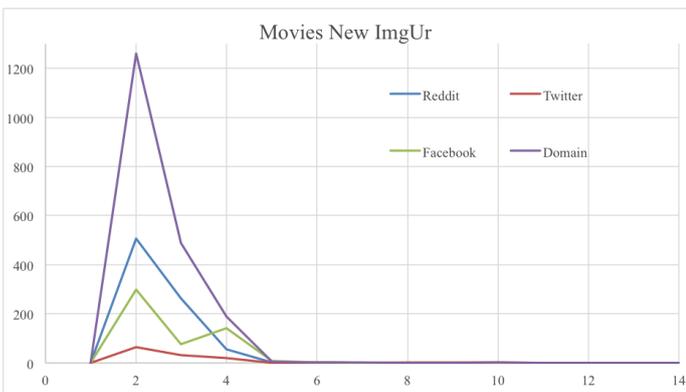
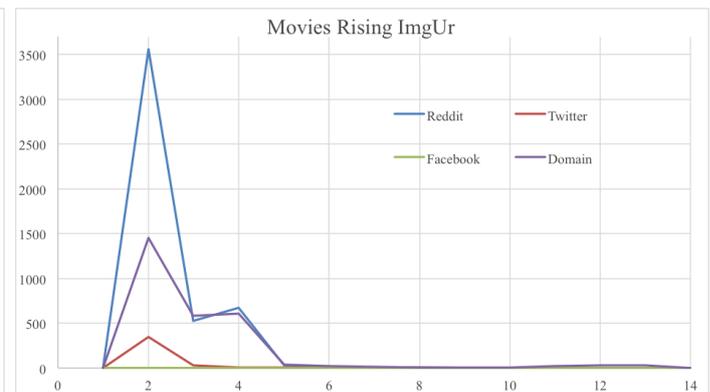

# PATTERN 3: Topic / Category / Domain: ImgUR

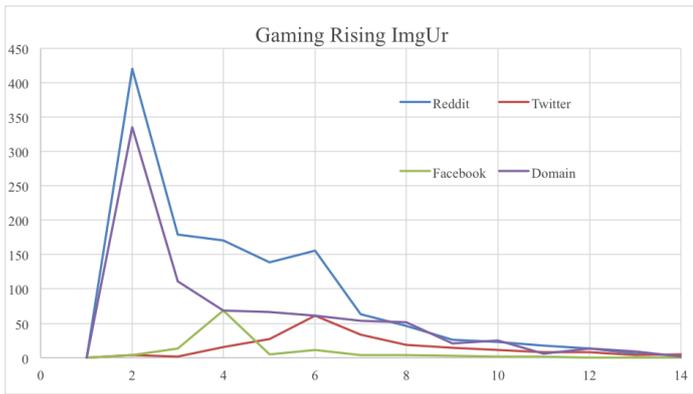
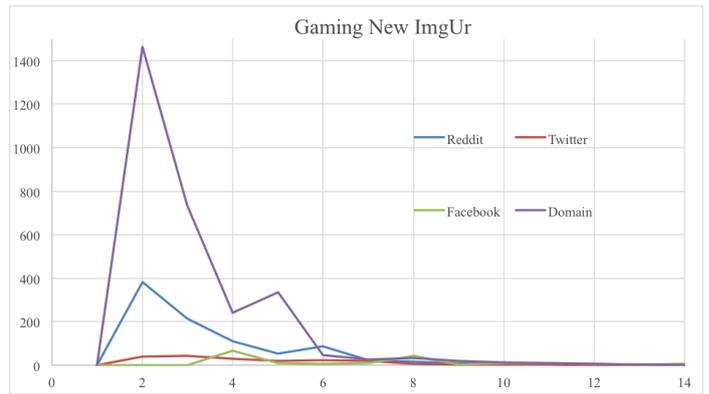

# PATTERN 4: Topic / Category / Domain: YouTube

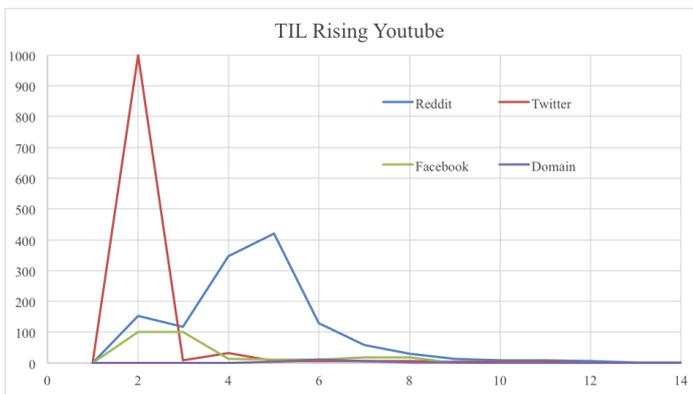
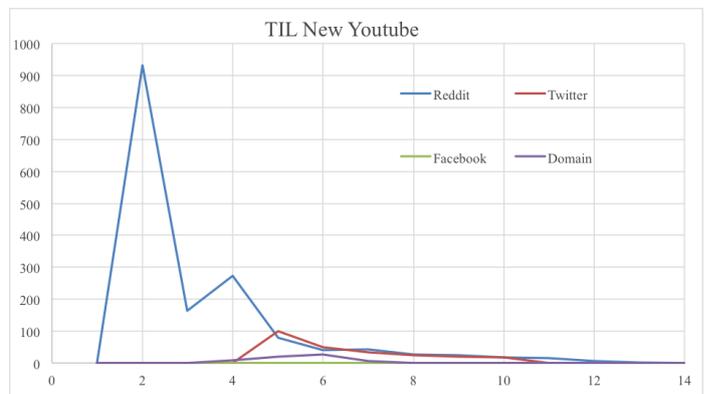

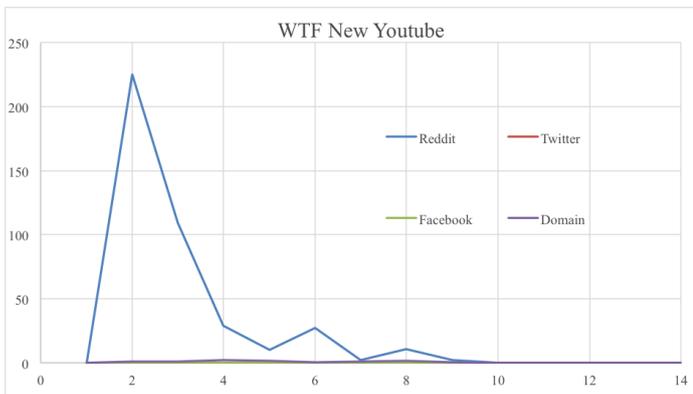
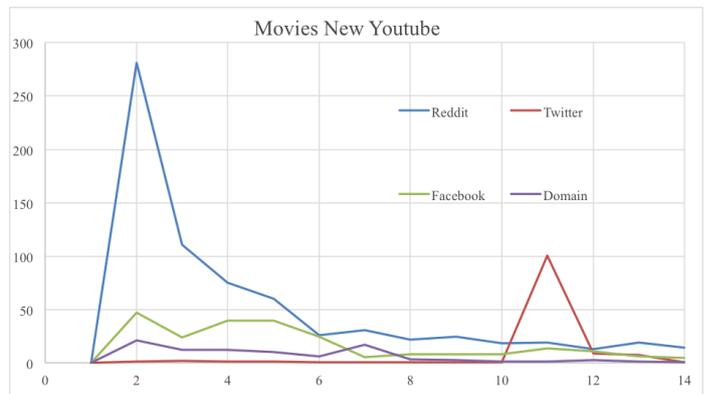

# PATTERN 5: Topic / Category / Domain: YouTube

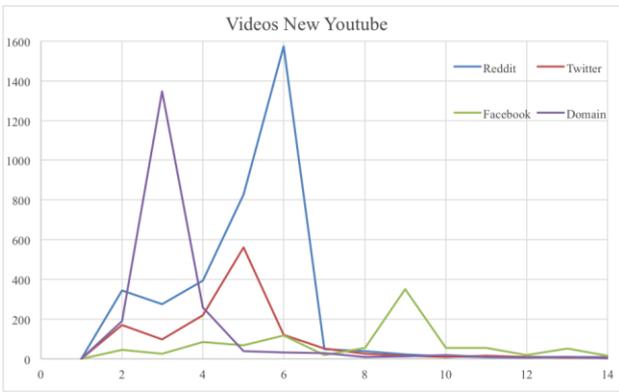
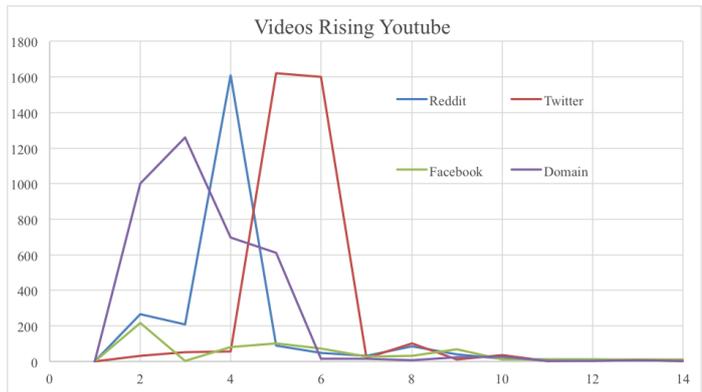
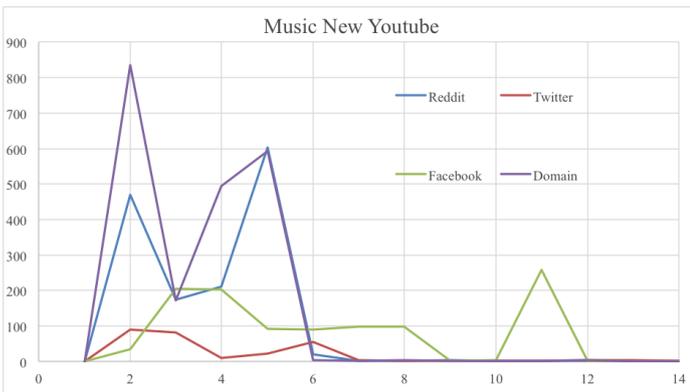
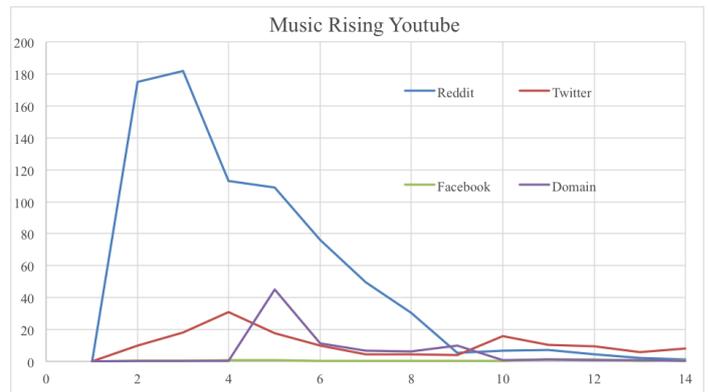
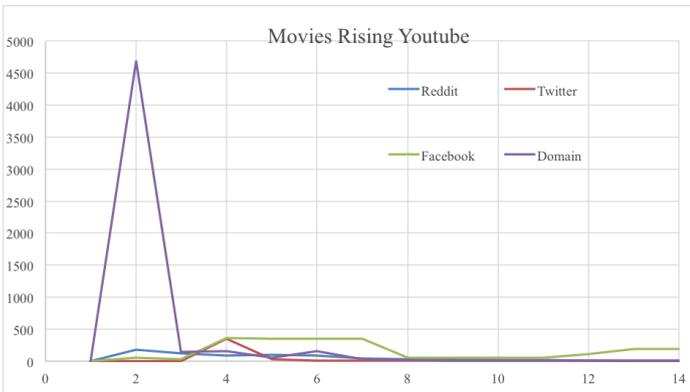